\documentclass[11pt]{article}
 \usepackage{titlesec}
\usepackage{hyperref}

\titleclass{\subsubsubsection}{straight}[\subsection]

\usepackage[utf8]{inputenc}
\usepackage{amsmath,amsfonts,amssymb,stackengine,graphicx}

\usepackage[utf8]{inputenc}
\newif\iffigs\figstrue

\usepackage[title]{appendix}
\usepackage{epsfig,latexsym}
\usepackage{hyperref}
\usepackage{amsmath}
\usepackage{verbatim}
\usepackage{color}
\usepackage{mathrsfs}
\usepackage{slashed}
\usepackage{amssymb}

\DeclareMathAlphabet{\mathpzc}{OT1}{pzc}{m}{it}

 \csname
@addtoreset\endcsname{equation}{section}

\def\gz0{\gamma^{0}}

 \def\det{{\rm det\,}}

\def\k{\kappa}

\def\m{\mu}

\def\beq{\begin{equation}}
\def\eeq{\end{equation}}
\def\bea{\begin{eqnarray}}
\def\eea{\end{eqnarray}}
\def\ba{\begin{array}}
\def\ea{\end{array}}
\def\bec{\begin{center}}
\def\ec{\end{center}}
\def\ba{\begin{align}}
\def\ena{\end{align}}

\def\12{\frac{1}{2}}

\newcounter{subsubsubsection}[subsubsection]
\renewcommand\thesubsubsubsection{\thesubsubsection.\arabic{subsubsubsection}}

\titleformat{\subsubsubsection}
  {\normalfont\normalsize\bfseries}{\thesubsubsubsection}{1em}{}
\titlespacing*{\subsubsubsection}
{0pt}{3.25ex plus 1ex minus .2ex}{1.5ex plus .2ex}

\makeatletter
\renewcommand\paragraph{\@startsection{paragraph}{5}{\z@}%
  {3.25ex \@plus1ex \@minus.2ex}%
  {-1em}%
  {\normalfont\normalsize\bfseries}}
\renewcommand\subparagraph{\@startsection{subparagraph}{6}{\parindent}%
  {3.25ex \@plus1ex \@minus .2ex}%
  {-1em}%
  {\normalfont\normalsize\bfseries}}
\def\toclevel@subsubsubsection{4}
\def\toclevel@paragraph{5}
\def\toclevel@paragraph{6}
\def\l@subsubsubsection{\@dottedtocline{4}{7em}{4em}}
\def\l@paragraph{\@dottedtocline{5}{10em}{5em}}
\def\l@subparagraph{\@dottedtocline{6}{14em}{6em}}
\makeatother

\begin{document}

\begin{flushright}
{\today}
\end{flushright}

\vspace{10pt}

\begin{center}


{\Large\sc Effective Orientifolds from Broken Supersymmetry}


\vspace{25pt}
{\sc J.~Mourad${}^{\; a}$  \ and \ A.~Sagnotti${}^{\; b}$\\[15pt]

${}^a$\sl\small APC, UMR 7164-CNRS, Universit\'e   Paris Cit\'e  \\
10 rue Alice Domon et L\'eonie Duquet \\75205 Paris Cedex 13 \ FRANCE
\\ e-mail: {\small \it
mourad@apc.univ-paris7.fr}\vspace{10
pt}

{${}^b$\sl\small
Scuola Normale Superiore and INFN\\
Piazza dei Cavalieri, 7\\ 56126 Pisa \ ITALY \\
e-mail: {\small \it sagnotti@sns.it}}\vspace{10pt}
}

\vspace{40pt} {\sc\large Abstract}\end{center}
\noindent

We recently proposed a class of type IIB vacua that yield, at low energies, four--dimensional Minkowski spaces with broken supersymmetry and a constant string coupling. They are compactifications with an internal five-torus bearing a five--form flux $\Phi$ and warp factors depending on a single coordinate. The breaking of supersymmetry occurs when the internal space includes a finite interval. A probe-brane analysis revealed a gravitational repulsion and a charge attraction of equal magnitude from the left end of the interval, {together with} a singularity at the other end. Here we complete the analysis revealing the presence, at one end, of an effective $O3$ of negative tension and positive five--form charge. We also determine the values of these quantities, showing that $T = -\, Q = \Phi$, and characterize the singularity present at the other end of the interval, which hosts an opposite charge. Finally, we discuss various forms of the gravity action in the presence of a boundary and identify a self--adjoint form for its fluctuations.

\vskip 12pt

\begin{center}
\emph{Invited contribution to the special issue of Journal of Physics A: Mathematical and Theoretical on
``Fields, Gravity, Strings and Beyond: In Memory of Stanley Deser''}
\end{center}

\setcounter{page}{1}

\pagebreak

\newpage
\tableofcontents
\newpage
\baselineskip=20pt
\section{\sc  Introduction and Summary}\label{sec:intro}

Different scenarios for supersymmetry breaking in String Theory~\cite{strings} have been explored over the years. The resulting pictures are captivating, but they all entail, in one way or another, strong back reactions on the vacuum, with the typical emergence of runaway potentials. These arise from quantum corrections in Scherk--Schwarz compactifications~\cite{ss_closed,ss_open}, and already at the (projective) disk level for the non--tachyonic ten--dimensional strings of~\cite{so1616,susy95,sugimoto}. 
The first model is a variant of the heterotic string, while the others are orientifolds~\cite{orientifolds}. Supersymmetry is absent in the first two, while it is non--linearly realized in the third~\cite{bsb_nonlinear}, which provides the simplest instance of ``brane supersymmetry breaking''~\cite{bsb}. 
The Dudas--Mourad vacua~\cite{dm_vacuum} provide, in all these cases, compactifications to lower--dimensional Minkowski spaces that are perturbatively stable~\cite{bms,selfadjoint}, notwithstanding the breaking of supersymmetry. These vacua have the interesting feature of including an internal interval, but in some regions the string coupling, and/or the space--time curvature, become unbounded. On the other hand, more conventional fluxed AdS vacua~\cite{gubsermitra,raucci_22}, where curvatures and string couplings are everywhere weak, typically host unstable modes~\cite{bms}.

This paper concerns a class of type IIB compactifications to four--dimensional Minkowski space with internal fluxes~\cite{ms21_2,ms22_1,ms23_3}, which also include an interval but avoid the emergence of regions where the string coupling becomes unbounded~\footnote{The strong curvatures present in these vacua can be confined to small portions of the internal space with suitable choices of their free parameters.}. In~\cite{ms22_1}, a probe brane was shown to experience a gravitational repulsion and a five--form charge attraction of equal magnitude near the left end of the interval, while a singularity whose features were less transparent revealed itself at the other end. The purpose of the present paper is to take a closer look at the endpoints, in order to unveil the key properties of the objects present there. The result will be especially neat for the left end, and the present analysis should also favor the comparison between the work of~\cite{ms21_1,ms21_2} and the current literature on ``dynamical cobordism'', some of which can be found in~\cite{dynamicalcobordism}.

The backgrounds of interest are characterized by a constant dilaton profile $\phi_0$, which we shall set to zero for brevity, and by metric and five-form profiles that depend on a single coordinate, $r$, and are given by
    \bea{}{}{}{}{}{}{}{}{}{}{}{}{}
ds^2 &=&  e^{\,2\,A(r)}\, dx^2 \ + \ e^{\,2\,B(r)}\, dr^2 \ + \ e^{\,2\,C(r)}\, dy^2 \nonumber \\
&=& \frac{\eta_{\mu\nu}\,dx^\mu\,dx^\nu}{\left[2\left|H\right|\rho\,\sinh\left(\frac{r}{\rho}\right)\right]^\frac{1}{2}} \,+\, \left[2 \left|H\right|\,\rho\,\sinh\left(\frac{r}{\rho}\right)\right]^\frac{1}{2} \left[e^{ \,- \, \frac{\sqrt{10}}{2\rho}\, r} \, dr^2 \ + \ e^{\,- \, \frac{\sqrt{10}}{10\rho}\, r} \, \left(d\,{y}^i\right)^2\right] \ , \nonumber \\
{{\cal H}_5^{(0)}} &=&
H\left\{ \frac{dx^0 \wedge ...\wedge dx^3\wedge dr}{\left[2\left|H\right|\,\rho\,\sinh\left(\frac{r}{\rho}\right)\right]^2} \ + \ dy^1 \wedge ... \wedge dy^5\right\} \ . \label{back_epos_fin2}
\eea
The $x^\mu$ are coordinates of a four--dimensional Minkowski space, and positive values of $r$ parametri\-ze the interior of the internal interval. The five $y^i$ coordinates have a finite range,
\beq{}{}{}{}{}{}{}{}{}{}{}{}{}{}{}{}{}{}{}{}{}{}{}{}{}{}{}{}{}
0 \ \leq \ y^i \ \leq \ 2\,\pi\,R \ ,
\eeq
and parametrize an internal torus, which for simplicity we take to be the direct product of five circles of radius $R$. {The other parameters that enter the background, $\rho$ and $H$, emerge as integration constants from the equations of the low--energy supergravity. The former characterizes the length of the internal interval parametrized by the variable $r$, while the latter clearly characterizes the five--form field strength. 
}

The contents of this paper are as follows. In Section~\ref{sec:bcs} we discuss boundary conditions at the singular ends of an interval for a toy scalar field theory, {relying on a conformal internal coordinate $z$,} with emphasis on the distinction between the first--order formulation and the self--adjoint, second--order one. In Section~\ref{sec:gen_rel_boundary} we adapt the Arnowitt--Deser--Misner (ADM) decomposition~\cite{ADM} to our setting and discuss the role of the York--Gibbons--Hawking term~\cite{ygh} in defining a first--order action for gravity. We also describe a self--adjoint form for its four--dimensional and internal traceless fluctuations. In Section~\ref{sec:background_or} we identify the opposite values of the tension and charge of an effective BPS $O_3$ orientifold that, as we had anticipated in~\cite{ms22_1}, lies at one end of the interval. {The result, }
\beq
T \ = \ - \ Q \ = \ - \ \frac{\Phi}{k_{10}^2} \ ,
\eeq
{is particularly simple: both quantities are proportional to the five--form flux on the internal torus}
\beq
\Phi \ = \ H \left( 2\,\pi\,R \right)^5 \ .
\eeq

{To the best of our knowledge, this is the first time that the effective emergence of a BPS object is revealed at the endpoint of a compactification. Clearly supersymmetry plays a role in our derivation, and the link between charge and tension stabilizes the latter. We also show that the singularity at the opposite end corresponds to an extended object with the expected opposite charge, but its characterization is admittedly less neat, since it involves a contribution proportional to the extrinsic curvature of that boundary and another (singular) tension--like contribution. Presumably, quantum corrections will play a role in determining the final form of these contributions.}
Appendix~\ref{app:background} collects some properties of the background that are useful in our derivations.

The ADM decomposition, which plays an important role in our considerations, ranks highly among the many important contributions that Stanley Deser gave to Theoretical Physics over the years. We are honored to contribute the present article to the volume dedicated to his memory.

\section{\sc Boundary Terms and Boundary Conditions} \label{sec:bcs}

Our background includes an interval of finite length, which can parametrized by the variable $r$ in eqs.~\eqref{back_epos_fin2} valued in the range $0 < r < \infty$, or alternatively by a conformal variable, related to $r$ according to
\beq
z \ = \ \int_0^r e^{B(\xi)-A(\xi)}\ d\xi \ ,
\eeq
which has also a finite span $0 < z < z_m$. When working in terms of $z$, the metric in eqs.~\eqref{back_epos_fin2} takes the form
\beq
ds^2 \ = \  e^{\,2\,A(z)}\left( dx^2 \ + \ dz^2\right) \ + \ e^{\,2\,C(z)}\, dy^2 \ , 
\eeq
and
\beq
z_m \ = \ \int_0^\infty dr'\ e^{B(r')-A(r')} 
\eeq
characterizes the range of the new variable.

One should demand that the action yield the equations of motion if it is stationary under arbitrary variations of all fields in the bulk. To this end, one must supplement it with boundary conditions to eliminate the boundary terms that accompany the field equations. Before addressing our problem, it is convenient to consider a toy model, a complex scalar field in a five--dimensional spacetime that also includes an interval parameterized by a real variable $z$,  with $0 < z < z_m$, and with the background metric
\beq
ds^2 \ = \ e^{2A(z)} \left( \eta_{\mu\nu} \ dx^\mu\ dx^\nu \ + \ dz^2 \right) \ ,
\eeq
where the exponential factor behaves as a power at both ends of the interval:
\beq
e^{2A} \ \sim \ z^{2\,\alpha_0} \ , \qquad e^{2A} \ \sim \ \left(z_m \,-\,z\right)^{2\,\alpha_m} \ .
\eeq
The standard presentation of the equation of motion
\beq
\frac{1}{\sqrt{\,-\,g}} \ \partial_M \left[\sqrt{\,-\,g} \ g^{MN} \right]\partial_N\, \phi \ = \ 0
\eeq
for four--dimensional modes of mass $m$ can be turned into the Schr\"odinger form
\beq
{\cal H}\,\Psi \ \equiv \ - \ \partial_z^2\,\Psi \ + \ V(z) \, \Psi  \ = \ m^2\, \Psi \label{2.4}
\eeq
by the field redefinition
\beq
\phi \ = \ \Psi \ e^{\,-\,\frac{3}{2}\, A} \ , \label{phipsi}
\eeq
and the resulting potential is
\beq
V(z) \ = \ \frac{9}{4}\ A_z^2 \ + \ \frac{3}{2}\ A_{zz} \ ,
\eeq
where $A_z$ and $A_{zz}$ indicate the first and second derivatives of $A(z)$. Note that close to the two ends the potential $V(z)$ behaves as
\beq
V(z) \ \sim \ \frac{\mu_0^2 \ - \ \frac{1}{4}}{z^2} \ , \qquad V(z) \ \sim \ \frac{\mu_m^2 \ - \ \frac{1}{4}}{\left(z_m\,-\,z\right)^2} \ ,
\eeq
where
\beq
\mu_0^2 \ = \ \frac{1}{4} \left(3\,\alpha_0 \ - \ 1\right)^2 \ , \qquad \mu_m^2 \ = \ \frac{1}{4} \left(3\,\alpha_m \ - \ 1\right)^2 \ .
\eeq

As discussed in~\cite{selfadjoint}, the Hamiltonian in eq.~\eqref{2.4} can be self--adjoint if it is supplemented with proper boundary conditions. These depend crucially on the behavior at the two ends, and specifically on whether $0 \leq \mu_{0,m} < 1$ or $\mu_{0,m} \geq 1$. In detail, for boundary conditions given independently at the ends, 
\begin{itemize}
    \item if $0 < \mu_{0} < 1$, the self--adjoint boundary conditions that are allowed at $z=0$ depend on a real parameter. This result reflects the possibility of allowing, at that end, the general limiting behavior available for $\Psi$, 
    \beq
\Psi \ \sim \ C_1 \ \left(\frac{z}{z_m}\right)^{\frac{1}{2} + \mu_0} \ + \ C_2 \ \left(\frac{z}{z_m}\right)^{\frac{1}{2} - \mu_0} \ , \label{2.9}
    \eeq   
    compatibly with the conditions that $\Psi$ and ${\cal H}\,\Psi$ be both in $L^2$. The possible self--adjoint extensions are in one--to--one correspondence with the different values of the ratio $\frac{C_1}{C_2}$. Similar considerations hold for the other end, if $0 < \mu_{m} < 1$,  with
    \beq
\Psi \ \sim \ C_3 \ \left(1\ - \ \frac{z}{z_m}\right)^{\frac{1}{2} + {\mu}_m} \ + \ C_4 \ \left(1 \ - \ \frac{z}{z_m}\right)^{\frac{1}{2} - {\mu}_m} \ , \label{2.91}
    \eeq 
    \item If $\mu_0=0$ or $\mu_m=0$, there are logarithmic contributions, and
        \bea
\Psi &\sim& C_1 \ \left(\frac{z}{z_m}\right)^{\frac{1}{2}}\ \log\left(\frac{z}{z_m}\right)\ + \ C_2 \ \left(\frac{z}{z_m}\right)^{\frac{1}{2}} \ , \nonumber \\
\Psi &\sim& C_3 \ \left(1\ - \ \frac{z}{z_m}\right)^{\frac{1}{2}}\ \log\left(1\ - \ \frac{z}{z_m}\right) \ + \ C_4 \ \left(1 \ - \ \frac{z}{z_m}\right)^{\frac{1}{2}}
\label{2.912} \ , \label{mu0m0}
    \eea
    and both types of limiting behaviors are allowed.
    \item In the complementary range $\mu_0 \geq 1$ there is a unique choice of self--adjoint boundary conditions, with $C_2=0$, since the other limiting behavior is incompatible with the $L^2$ condition, and similarly at the other end one must choose $C_4=0$ if $\mu_m \geq 1$.
\end{itemize}

Let us now consider the action for a complex scalar field $\phi$ in the second--order form, while focusing on four--dimensional mass eigenstates, which reads
\beq
{\cal S} \ = \ \int_{\cal M} \ d^4 x \ dz \ \phi^\star \ \partial_M\,\Big( \sqrt{-g}\ g^{MN} \partial_N \ \phi\Big)  \ . \label{2.11}
\eeq
Performing the field redefinition~\eqref{phipsi}, this action takes the form
\beq
{\cal S} \ = \ \int_{\cal M} \ d^4 x \ dz \ \Psi^\star \left(  {\cal H} \ - \ m^2 \right) \Psi  \ , \label{2.112}
\eeq
and its variation reads
\bea
\delta\,{\cal S} &=& \lim_{z^\star\to 0,\, Z^\star \to z_m} \ 2\ \int_{{\cal M}^\star} \ d^4 x \ dz \ \delta \Psi^\star \left(  {\cal H} \ - \ m^2 \right) \Psi \nonumber \\ &-& \ \lim_{z^\star\to 0,\, Z^\star \to z_m} \ \left. \int \ d^4 x \left( \Psi^\star \ \partial_z \ \delta\,\Psi \ - \  \partial_z\,\Psi^\star \ \delta\,\Psi \right) \right|_{z=z^\star}^{z=Z^\star} \ . \label{sadjpsi}
\eea

It is important to stress that the field $\Psi$ in eq.~\eqref{2.112} and ${\cal H}\,\Psi$ should be both in $L^2$, so that $\Psi$ should behave as in eqs.~\eqref{2.9} or \eqref{2.91} near the ends of the interval. Moreover, when two coefficients are allowed in the limiting behavior, the variation $\delta\,\Psi$ is computed for a fixed value of their ratio so that, for example, for the behavior in eq.~\eqref{2.9}
\beq
\delta\,\Psi \ \sim \ \delta\,C_2 \left[  \frac{C_1}{C_2} \ \left(\frac{z}{z_m}\right)^{\frac{1}{2} + \mu_0} \ + \ \left(\frac{z}{z_m}\right)^{\frac{1}{2} - \mu_0} \right] \ , \label{deltapsi1}
\eeq
if the ratio $\frac{C_1}{C_2}$ is finite, and otherwise
\beq
\delta\,\Psi \ \sim \ \delta\,C_1 \ \left(\frac{z}{z_m}\right)^{\frac{1}{2} + \mu_0} \ .  \label{deltapsi2}
\eeq
The preceding conditions grant the vanishing of the boundary term in eq.~\eqref{sadjpsi}, and the consequent recovery of the Schr\"odinger--like equation~\eqref{2.4} from the action principle.

Summarizing, when singularities are present at the ends of the interval, the boundary conditions are not defined directly in terms of $\Psi$ and its derivative. Rather, they are defined in terms of the ratios $\frac{C_1}{C_2}$ and $\frac{C_3}{C_4}$ of the coefficients characterizing their limiting behavior, which are well defined despite the fact that this is generally singular.
Note that, to this end, we displaced the two endpoints to $z^\star$ and $Z^\star$, slightly away from the singular ends at $z=0$ and $z=z_m$ and toward the interior of the interval. The resulting regularized manifold is denoted by ${\cal M}^\star$, and we shall follow the same procedure in the following section.  

The action~\eqref{2.112} is equivalent to the first--order form
\beq
{\cal S} \ = \  - \ \int_{{\cal M}^\star} d^4 x\ dz \sqrt{- g}\ \partial^{M}\,\phi^\star \ \partial_M\, \phi \ + \  \lim_{z^\star\to 0,\, Z^\star \to z_m} \ \left. \int d^4 x \ \sqrt{-g} \ \phi^\star \, g^{z N}\, \partial_N\, \phi\ \right|_{z=z^\star}^{z=Z^\star} \ ,
\eeq
or, in terms of $\Psi$, to
\beq
{\cal S} \ = \  \int_{{\cal M}^\star} d^4 x \ dz \Big[ - \ \left|\partial_z\,\Psi\right|^2 \ \ + \ \left(m^2\ - \ V(z)\right) \left|\Psi\right|^2 \Big] \ + \ \lim_{z^\star\to 0,\, Z^\star \to z_m} \ \left. \int d^4 x \ \Psi^\star \, \partial_z\, \Psi\ \right|_{z=z^\star}^{z=Z^\star} \ . \label{2.14}
\eeq
The second contribution to this expression is singular when $C_2$ or $C_4$ do not vanish, and gives rise to divergent boundary terms proportional to $\left|C_2\right|^2$ and $\left|C_4\right|^2$. These contributions are compensated by other singular portions of the first term, while eq.~\eqref{2.112} contains no singular terms, since by assumption ${\cal H}\,\Psi$ is in $L^2$. Consequently, the standard practice of removing the boundary term in eq.~\eqref{2.14}, while also insisting on the same set of $\Psi$ eigenfunctions, would lead to divergent contributions proportional to $\left|C_2\right|^2$ and $\left|C_4\right|^2$. The first--order action without the boundary term is thus equivalent to the Schr\"odinger--like form~\eqref{2.112} only for boundary conditions whereby $C_2$ and $C_4$ vanish. These considerations extend to other bosonic fluctuations, and in particular to the gravitational field, to which we can now turn. 

\section{\sc General Relativity and Boundary Terms} \label{sec:gen_rel_boundary}

As was the case for the scalar field, the Einstein--Hilbert action can be linked to a first--order form, but neither of them is the analog of the self--adjoint action~\eqref{2.112}, whose variation leads to the Schr\"odinger--like equation
\beq
\left({\cal H} \ - \ m^2\right)\Psi \ = \ 0 \ ,
\eeq
once it is supplemented by self--adjoint boundary conditions. 

Let us see this in detail, recalling, to begin with, that the Einstein--Hilbert action ${\cal S}_{EH}$ can be related to a first--order form, ${\cal S}_G$, by the addition of a York--Gibbons--Hawking term~\cite{ygh} at the spatial boundary. This boundary term is conveniently formulated in terms of an ADM--like decomposition~\cite{ADM} that singles out constant-$z$ hypersurfaces. The decomposition rests on a nine--dimensional symmetric tensor $\tilde{g}_{mn}$, which is the metric on constant-$z$ hypersurfaces, on a nine--dimensional ``shift'' vector ${\cal N}_m$ and on a ``lapse'' function ${\cal N}$. A ten--dimensional label $M$ thus splits into $z$ and a nine--dimensional label $m$, and
\beq{}{}{}{}{}{}{}{}{}{}{}{}{}{}{}{}{}{}{}{}{}{}{}{}{}{}{}{}{}{}{}{}{}{}{}{}{}{}{}{}{}{}{}{}{}{}{}{}{}{}{}{}{}{}{}{}{}{}{}{}{}{}{}{}{}{}{}{}{}{}{}{}{}
\tilde{g}_{mn} \ = \ g_{mn} \ , \qquad {\cal N}_m \ = \ g_{mz}  \ , \qquad {\cal N}^2 \ + \ {\cal N}^m\,{\cal N}_m \ = \ g_{zz} \ . \label{defs}
\eeq
Nine--dimensional indices are raised and lowered with the nine--dimensional metric $\tilde{g}$, whose associated covariant derivatives and scalar curvature we denote by $\widetilde{D}_m$ and $\widetilde{R}$, and another important ingredient is the extrinsic curvature of the boundary,
\beq
{\cal K}_{mn} \ = \ \frac{1}{2\,{\cal N}}\left(\partial_z\,\tilde{g}_{mn} \ - \ \widetilde{D}_{(m}\,{\cal N}_{n)} \right) \ . \label{extrinsic}
\eeq
The Einstein--Hilbert action then decomposes as
\beq
{\cal S}_{EH}\ = \ {\cal S}_{G}\ - \ \lim_{z^\star\to 0,\, Z^\star \to z_m} \ \frac{1}{2\, k_{10}^2}  \left. \int d^9 \, x\ \sqrt{-\tilde{g}} \ {\cal K} \ \right|_{z^\star}^{Z^\star} \label{SehSg}
\eeq
where 
\beq
{\cal K} \ = \ \tilde{g}^{mn}
\ {\cal K}_{mn} \ ,
\eeq
and ${\cal S}_G$, the first--order action for the gravitational field, reads
\beq{}{}{}{}{}{}{}{}{}{}{}{}{}{}{}{}{}{}{}{}{}{}{}{}{}{}{}{}{}{}{}{}{}{}{}{}{}{}{}{}{}{}{}{}{}{}{}{}{}{}{}{}{}{}{}{}{}{}{}{}{}{}{}{}{}{}{}{}{}{}{}
{\cal S}_{G} \ = \ \frac{1}{2\,k_{10}^2}\int_{{\cal M}^\star} \ d^9 \,x \, dz \,\sqrt{-\tilde{g}}\,{\cal N} \left[ \widetilde{R} \ + \ {\cal K}_{mn}\,{\cal K}_{pq}\left(\tilde{g}^{mn}\,\tilde{g}^{pq} \ - \ \tilde{g}^{mp}\,\tilde{g}^{nq} \right)\right]  \label{EYGH0} \ .
\eeq

By construction, the variation of ${\cal S}_G$ contains a boundary term that originates solely from the last contributions involving the extrinsic curvature, and
\bea
\delta\,{\cal S}_G &=& \frac{1}{2\,k_{10}^2} \,\int_{{\cal M}^\star} d^{10}\,x \ \sqrt{-\tilde{g}}\ {\cal N} \ \delta\,g^{MN}\ G_{MN}  \nonumber \\
&+& \lim_{z^\star\to 0,\, Z^\star \to z_m} \ \left. \frac{1}{2\,k_{10}^2} \ \int \ d^9 \,x\, \sqrt{- \tilde{g}}\ \delta \tilde{g}^{mn} \, \Big( {\cal K}_{mn} \ - \ \tilde{g}_{mn}\,{\cal K} \Big)\right|_{z=z^\star}^{z=Z^\star} \label{2.5o}
\eea
does not contain any terms involving $\partial_z\,\delta\,\tilde{g}^{mn}$ at the boundary. This property makes ${\cal S}_G$ the counterpart of the first--order scalar action discussed in the previous section, and this very fact motivated the modification of the Einstein--Hilbert action proposed in~\cite{ygh}. In contrast, the self--adjoint formulation for the scalar field described in the previous section does accommodate terms containing both $\partial_z\,\delta\,\Psi$ and $\delta\,\Psi$, as can be seen in eq.\eqref{sadjpsi}. However, the two quantities are not independent. In fact, in view of eqs.~\eqref{deltapsi1} and \eqref{deltapsi2} the limiting behavior of the scalar field near the boundary relates $\delta\,\Psi$ and $\partial_z\,\delta\,\Psi$ there, according to
\beq
\partial_z\,\delta\,\Psi \ = \ \frac{f_0(z)}{z} \ \delta\,\Psi \ , \qquad \partial_z\,\delta\,\Psi \ = \ \frac{f_m(z)}{z_m - z} \ \delta\,\Psi  \ .\label{dzpsi}
\eeq
The two functions $f_{0}(z)$ and $f_{m}(z)$ approach constant values near the boundary, which depend on the given choice of self--adjoint boundary conditions, and more precisely on the indicial exponents characterizing the limiting behavior allowed by them. As we shall see in~\cite{ms23_3}, $\mu_0\neq 0$ in all bosonic mode sectors of the background~\eqref{back_epos_fin2}, and consequently at a point $z^\star$ close to $z=0$
\beq
f_0(z^\star) \ = \ \frac{C_1 \left(\frac{1}{2} \,+\,\mu_0\right) \left(\frac{z^\star}{z_m}\right)^{2\,\mu_0} \ + \  C_2 \left(\frac{1}{2} \,-\,\mu_0\right)}{C_1 \, \left(\frac{z^\star}{z_m}\right)^{2\,\m_0} \ + \  C_2 } \ ,
\eeq
which approaches $\frac{1}{2} \,+\,\mu_0$ if $C_2=0$, and $\frac{1}{2} \,-\,\mu_0$ otherwise. On the other hand, if $\mu_m \neq 0$,  at a point $Z^\star$ close to $z=z_m$
\beq
f_m(Z^\star) \ = \ - \ \frac{C_3 \left(\frac{1}{2} \,+\,\mu_m\right) \left(1 \ - \ \frac{Z^\star}{z_m}\right)^{2\,\mu_m} \ + \  C_4 \left(\frac{1}{2} \,-\,\mu_m\right)}{C_3 \, \left(1 \ - \ \frac{Z^\star}{z_m}\right)^{2\,\mu_m} \ + \  C_4 } \ ,
\eeq
which approaches $- \left(\frac{1}{2} \,+\,\mu_m\right)$ if $C_4=0$, and $- \left( \frac{1}{2} \,-\,\mu_m\right)$ otherwise. However, if $\mu_m=0$, one should start from eqs.~\eqref{mu0m0}
and then
\beq
f_m(Z^\star) \ = \ - \ \frac{1}{2}\ \frac{C_3 \,\log\left(1 \,-\, \frac{Z^\star}{z_m}\right) \ + \ 2 \ C_3 \ + \ C_4}{C_3\,\log\left(1 \,-\, \frac{Z^\star}{z_m}\right) \ + \ C_4} \ ,
\eeq
which approaches $-\,\frac{1}{2}$ in all cases.

Einstein's theory is highly non linear, and thus far more complicated than the scalar toy model, but the analogy is nonetheless very useful in the study of its linear fluctuations around the background~\eqref{back_epos_fin2}. As we saw in~\cite{selfadjoint}, after proper field redefinitions the bosonic fluctuations of the Dudas--Mourad vacua~\cite{dm_vacuum} can be linked to Schr\"odinger--like systems with double--pole singularities at the ends. This type of analysis is extended to the different bosonic sectors of the more complicated background~\eqref{back_epos_fin2} in~\cite{ms23_3}, where we show, in particular, that two Schr\"odinger fields $\Psi_{1}(z)$ and $\Psi_{2}(z)$ with identical potentials, which can be defined according to
\beq
h_{\mu\nu}(x,z) \ = \  e^\frac{A\,-\,5 C}{2} \ h_{\mu\nu}(x) \, \Psi_1(z) \ , \qquad
h_{ij}(x,z) \ = \  e^{\,-\,\frac{3A \,+\, C}{2}} \ h_{ij}(x) \, \Psi_2(z) \ , \label{redefs}
\eeq
describe both traceless spin-2 $h_{\mu\nu}$ and traceless spin-0 $h_{ij}$ fluctuations. The identical Schr\"odinger potentials for $\Psi_{1}(z)$ and $\Psi_{2}(z)$ have the limiting behavior discussed in the previous section, with $\mu_0=\frac{1}{3}$ and $\mu_m=0$. Making use of eqs.~\eqref{dzpsi}, one can thus obtain
\bea
\partial_z\,\delta\,h_{\mu\nu} &=& \frac{1}{z^\star} \left[ \frac{z^\star}{2}\left(A_z \,-\, 5\,C_z\right) \ + \ f_0(z^\star) \right]\delta\,h_{\mu\nu} \ , \nonumber \\
\partial_z\,\delta\,h_{ij} &=& \frac{1}{z^\star}  \left[ - \ \frac{z^\star}{2}\left(3\,A_z \,+\,C_z\right) \ + \ f_0(z^\star) \right] \delta\,h_{ij} \ ,
\eea
and taking the results in Appendix~\ref{app:background} into account, one can conclude that
\bea
\partial_z\,\delta\,h_{\mu\nu} &=& \pm \ \frac{\mu_0}{z^\star} \ \delta\,h_{\mu\nu}  \ = \ \pm \ \frac{1}{3\,z^\star}\  \delta\,h_{\mu\nu}, \nonumber \\
\partial_z\,\delta\,h_{ij} &=& \frac{\frac{2}{3} \ \pm \ \mu_0}{z^\star} \  \delta\,h_{ij} \ = \  \frac{2 \,\pm\,1}{3\,z^\star} \ \delta\,h_{ij}\ , \label{leftend}
\eea
where the upper sign applies if $C_2=0$.
In a similar fashion, near the other end of the interval, 
\bea
\partial_z\,\delta\,h_{\mu\nu} \!\!&=&\!\! \left[ \frac{A_z \,-\, 5\,C_z}{2} \ + \ \frac{f_m(Z^\star)}{z_m \ - \ Z^\star} \right]\delta\,h_{\mu\nu} \, = \, - \ \frac{2 \ + \ \sqrt{10}}{3\left(z_m \ - \ Z^\star\right)} \ \delta\,h_{\mu\nu}\ = \ 2 A_z\,\delta\,h_{\mu\nu} \,, \nonumber \\
\partial_z\,\delta\,h_{ij} \!\!&=& \!\!\left[ - \ \frac{3\,A_z \,+\,C_z}{2} \ + \ \frac{f_m(Z^\star)}{z_m \ - \ Z^\star} \right] \delta\,h_{ij} \, = \, \frac{\sqrt{10}}{5\left(z_m \ - \ Z^\star\right)} \ \delta\,h_{ij} \ = \  2\,C_z\,\delta\,h_{ij} \,. \label{rightend}
\eea
Note that the same relations linking the limiting behaviors of the fluctuations and their derivatives to $A_z$ and $C_z$ hold, at the left end, for the lower sign choices in eqs.~\eqref{leftend}, which apply whenever $C_2\neq 0$.

One can also identify a counterpart of the action~\eqref{2.112} for the gravitational field.  Confining initially the attention to the spin--2 variation $\delta\,g_{\mu\nu}$ alone, the second--order action is
\bea
{\cal S}_{sa} &=&   {\cal S}_G \ - \ \lim_{z^\star\to 0,\, Z^\star \to z_m} \left.  \frac{1}{4\,k_{10}^2}\  \left[ \int d^9 \, x\ \sqrt{-\tilde{g}} \ \Big( {\cal K} \ + \ \Lambda(z) \Big) \right]  \ \right|_{z=z^\star}^{z=Z^\star} \nonumber \\
&+& \lim_{z^\star\to 0,\, Z^\star \to z_m} \left. \frac{c(z)}{k_{10}^2} \ \int \epsilon_{\alpha_1 \ldots \alpha_4}\ e^{\alpha_1}\,\wedge\,\cdots\,e^{\alpha_4} \,\wedge \, {\cal H}_5 \right|_{z=z^\star}^{z=Z^\star} \label{SaSg}
\ ,
\eea
where $\Lambda(Z^\star)$ and $\Lambda(z^\star)$ are a pair of nine--dimensional cosmological constants, while the last term is pair of tension--like contributions, all localized at the two ends. In the background, the second pair of terms is equivalent to
\beq
\lim_{z^\star\to 0,\, Z^\star \to z_m} \left. \frac{H\  c(z)}{k_{10}^2} \ \int d^9\,x \ \sqrt{-\,\det\,\tilde{g}_{\mu\nu}} \right|_{z=z^\star}^{z=Z^\star}
\eeq

Using eq.~\eqref{2.5o}, one can now show that the full variation reads
\bea
&& \delta\,{\cal S}_{sa} = \frac{1}{2\,k_{10}^2} \,\int_{{\cal M}^\star} d^{10}\,x \ \sqrt{-\tilde{g}}\ {\cal N} \ \delta\,g^{MN}\ G_{MN} \nonumber \\
&&+ \lim_{z^\star\to 0,\, Z^\star \to z_m}  \frac{1}{4\,k_{10}^2} \ \left.  \int d^9 x \ \sqrt{-\tilde{g}} \left( \delta\,\tilde{g}^{mn} \ {\cal K}_{mn} \ - \ \tilde{g}^{mn}\, \delta\,{\cal K}_{mn} \ - \ \frac{3}{2}\ \delta\,\tilde{g}^{mn} \, \tilde{g}_{mn}\, {\cal K} \right) \ \right|_{z=z^\star}^{z=Z^\star}  \nonumber \\
&& + \lim_{z^\star\to 0,\, Z^\star \to z_m}  \frac{1}{8\,k_{10}^2} \left. \ \int d^9 x \ \sqrt{-\tilde{g}} \ \Lambda(z) \ \delta\,\tilde{g}^{mn} \ \tilde{g}_{mn}  \ \right|_{z=z^\star}^{z=Z^\star} \nonumber \\
&-& \lim_{z^\star\to 0,\, Z^\star \to z_m} \left. \frac{H\ c(z)}{2\ k_{10}^2} \ \int d^9\,x \ \sqrt{-\,\det\,g_{\mu\nu}} \ \delta\,\tilde{g}^{\mu\nu}\ \tilde{g}_{\mu\nu} \right|_{z=z^\star}^{z=Z^\star} \ . \label{delta_S_hmn}
\eea
For traceless and divergence--free $\delta\,g_{\m\nu}$ and $\delta\,g_{ij}$ the first two terms in the second line recover precisely the structure that emerged for the scalar field in eq.~\eqref{sadjpsi}. The correspondence becomes manifest performing the separation of variables and the first redefinition in eq.~\eqref{redefs}, after which the quadratic contributions around the background~\eqref{back_epos_fin2} become
\bea
\delta\,{\cal S}_{sa} &=&  \frac{1}{4\,k_{10}^2} \ \int_{{\cal M}^\star} d^9 x \ h_{\alpha\beta}(x)\,h^{\alpha\beta}(x) \ \delta\,\Psi_1 \Big(- \ \partial_z^2 \ + \ V(z) \ - \ m^2\Big)\Psi_1 \label{sa_variation} \\
&+&  \frac{1}{4\,k_{10}^2} \ \int_{{\cal M}^\star} d^9 x \ h_{ij}(x)\,h^{ij}(x) \ \delta\,\Psi_2 \Big(- \ \partial_z^2 \ + \ V(z) \ - \ m^2\Big)\Psi_2 \nonumber \\
&+& \lim_{z^\star\to 0,\, Z^\star \to z_m} \frac{1}{8 k_{10}^2} \left.  \int d^9 x \ h_{\alpha\beta}(x) \ h^{\alpha\beta}(x) \ \Big( \Psi_1 \, \partial_z\, \delta\,\Psi_1 \ - \ \delta\,\Psi_1 \ \partial_z\, \Psi_1 \Big) \ \right|_{z=z^\star}^{z=Z^\star} \nonumber \\
&+& \lim_{z^\star\to 0,\, Z^\star \to z_m} \frac{1}{8 k_{10}^2} \left.  \int d^9 x \ h_{ij}(x) \ h^{ij}(x) \ \Big( \Psi_2 \, \partial_z\, \delta\,\Psi_2 \ - \ \delta\,\Psi_2 \ \partial_z\, \Psi_2 \Big) \ \right|_{z=z^\star}^{z=Z^\star} \nonumber \\
&+& \lim_{z^\star\to 0,\, Z^\star \to z_m} \frac{1}{8 k_{10}^2}   \int d^9 x  \ h_{\alpha\beta}(x) \ h^{\alpha\beta}(x)  \nonumber \\
&\times& \left. \left[ 11 A_z \,+\, 20 C_z\,-\,e^A\left( \Lambda(z)\,+\,4\, H \, c(z) \ e^{-5C}\right)\right] \delta\,\Psi_1\,\Psi_1 \ \right|_{z=z^\star}^{z=Z^\star} \nonumber \\
&+& \lim_{z^\star\to 0,\, Z^\star \to z_m} \frac{1}{8 k_{10}^2} \left.  \int d^9 x  \ h_{ij}(x) \ h^{ij}(x) \ \left(15 A_z \,+\, 16 C_z \, - \, e^A\, \Lambda(z)\right)\, \delta\,\Psi_2 \, \Psi_2  \ \right|_{z=z^\star}^{z=Z^\star} \nonumber
,
\eea
where for brevity we are setting $h_{\alpha i}=0$ and we have used the four--dimensional mass--shell conditions
\beq
\Box\,h_{\alpha\beta} \ = \ m^2\, h_{\alpha\beta} \ , \qquad \Box\,h_{ij} \ = \ m^2\, h_{ij} 
\eeq
Using the results in the Appendix, one can see that the resulting potential for $\Psi_1$
\beq
V(z) \ = \ \frac{1}{4} \left(3 A_z \ + \ 5 C_z\right)^2 \ + \ \frac{1}{2} \left( 3 A_{zz} \ + 5 C_{zz}\right) 
\eeq
has double poles at the ends of the interval, with $\mu_0=\frac{1}{3}$ and $\mu_m=0$.

The two nine--dimensional cosmological terms $\Lambda(Z^\star)$ and $\Lambda(z^\star)$, and the four--dimensional one proportional to $c(z)$, can be chosen so that the last two lines in eq.~\eqref{sa_variation} vanish identically,
and
\beq
\Lambda(z) \ = \ e^{-A}\left(15\,A_z \,+\, 16\,C_z\right) \ , \qquad  H\,c(z) \ = \ e^{-A + 5C} \left(A_z \,+\,C_z\right) \ .
\eeq

At the lower end $z^\star$
\beq
\Lambda(z^\star) \ = \ \frac{\left(3\,H\right)^\frac{1}{6}}{6\, \left({z^\star}\right)^\frac{5}{6}} \ , \qquad  H\,c(z^\star) \ = \ 0 \ .
\eeq
while at the upper end $z=Z^\star$
\bea
\Lambda(Z^\star) &=& - \ \frac{50 \,+\,\sqrt{10}}{15}  \left(\frac{h}{2}\right)^\frac{1}{4} \left[\frac{\sqrt{5}-\sqrt{2}}{2} \left(\frac{z_m-Z^\star}{z_0}\right)\right]^{-\,\frac{\sqrt{10}\,+\,2}{6}}  \ , \nonumber \\
H\,c(Z^\star) &=&  \frac{5 \,+\,\sqrt{10}}{15} \  \left(\frac{h}{2}\right)^\frac{3}{2} \left[\frac{\sqrt{5}-\sqrt{2}}{2 \, z_0}\right]^{-\,\frac{5 \sqrt{10}\,+\,4}{12}}  \left({z_m-Z^\star}\right)^{\,-\,\frac{5 \sqrt{10}\,+\,16}{12}} .
\eea
Let us stress again the two main limitations of this analysis. First of all, it is confined to linearized perturbations, and moreover it leaves out lower--spin terms that receive contributions from the matter. Still, the self--adjoint form that was discussed in the previous section is recovered both for the spin--two perturbations described by $h_{\mu\nu}$ and for the traceless scalar ones described by $h_{ij}$. In the same spirit, the variation of the action vanishes on the background if the variations $\delta\,\tilde{g}^{\mu\nu}$ and $\delta\,\tilde{g}^{ij}$ are confined to their traceless portions.

Note also that the link in eq.~\eqref{SehSg} between ${\cal S}_G$ and the Einstein--Hilbert action ${\cal S}_{EH}$ implies that its variation is
\bea
\delta\,{\cal S}_{EH} &=& \frac{1}{2\,k_{10}^2} \int_{{\cal M}^\star} d^{10}\,x \ \sqrt{-\tilde{g}}\,{\cal N} \ \delta\,g^{MN}\ G_{MN}  \nonumber \\
&-& \lim_{z^\star\to 0,\, Z^\star \to z_m} \ \left.\frac{1}{2\,k_{10}^2} \ \int \ d^9 \,x\, \ \sqrt{- \tilde{g}}\left(  \frac{1}{2}\  \delta \tilde{g}^{mn} \, g_{mn}\, {\cal K}  \ + \ \tilde{g}^{mn}\, \delta{\cal K}_{mn} \right)  \right|_{z=z^\star}^{z=Z^\star} \ . \label{2.5}
\eea
Consequently, the Einstein--Hilbert action ${\cal S}_{EH}$ is somehow intermediate between the first--order and second--order actions for gravity fluctuations. 

Let us conclude this section by deducing from eq.~\eqref{delta_S_hmn} some properties of the boundary fields that can be present in this case, along the lines of what did in detail in~\cite{ms23_3} for the type--IIB two-forms. To begin with, when the metric field is varied in the bulk, the $r$ derivatives present in
\beq
\delta\,{g}_{m r} \ = \ \nabla_m\,\xi_r \,+\, \nabla_r\,\xi_m \ , \qquad \delta\,{g}_{r r} \ = \ 2\, \nabla_r\,\xi_r 
\eeq
give rise to the boundary terms in
\beq
\delta_\xi\,{\cal S}_{EH} \ = \  \frac{1}{k_{10}^2} \int_{\partial\,{\cal M}^\star} d^{9}\,x \ \sqrt{-\tilde{g}}\,{\cal N} \left[\xi_m\ G^{m r} \ + \ \xi_r\ G^{r r} \right] \ .
\eeq
These contributions can be canceled adding 
\beq
\Delta\,{\cal S} \ = \ \frac{1}{k_{10}^2} \int_{\partial\,{\cal M}^\star} d^{9}\,x \ \sqrt{-\tilde{g}}\,{\cal N} \left[A_m\ G^{m r} \ + \ A_r\ G^{r r} \right] \ , \label{DELTASA}
\eeq
where the St\"uckelberg fields, a real vector ${A}{}^{\,m}$ and a real scalar ${A}{}^{\,r}$, transform as
\beq
\delta\,{A}{}_{\,m} \ = \ - \ \xi_m \ , \qquad \delta\,{A}{}_{\,r} \ = \ - \ \xi_r \ .
\eeq

There is in principle a subtlety with $A_m$, which is a vector and should satisfy a gauge invariant equation of motion around the background of eqs.~\eqref{back_epos_fin2}. In analogy with what was done for the type--IIB two-forms in~\cite{ms23_3}, the kinetic operator for $A_m$ can be deduced from the variation $\delta\,g_{m r}$, when this is expressed in terms of the extrinsic curvature. From the bulk term ${\cal S}_G$ in the ADM decomposition of eq.~\eqref{EYGH0}, one can deduce the variation
\beq
\delta\,{\cal S}_G \ = \ \frac{1}{k_{10}^2}\ \int d^{10}\,x \ \sqrt{-\,\tilde{g}} \ \delta N_n \ \widetilde{D}{}_m \left[ K^{mn} \ - \ \tilde{g}{}^{mn} K\right] \ ,
\eeq
and the comparison with the corresponding expression in the standard ten--dimensional Einstein--Hilbert form links the mixed components $G^{m r}$ of the Einstein tensor to the extrinsic curvature, according to
\beq
G^{n r} \ = \ - \  \frac{1}{\cal N} \ \widetilde{D}{}_m \left[ K^{mn} \ - \ \tilde{g}{}^{mn} \,K \right] \ .
\eeq
Consequently the term in the action~\eqref{DELTASA} involving $A_m$ takes the form
\beq
\Delta\,{\cal S} \ = \ - \ \frac{1}{k_{10}^2}\ \int_{\partial\,{\cal M}^\star} d^9 \,x \ \sqrt{-\,\tilde{g}}\  A_m \ \widetilde{D}{}_m \left[ K^{mn} \ - \ \tilde{g}{}^{mn} \,K \right] \ ,
\eeq
and one can now vary $N_m$ in this expression, making use of eq.~\eqref{extrinsic}. After a partial integration one is thus led to
\beq
\delta\left(\Delta\,{\cal S}\right) \ = \ - \ \frac{1}{2\,k_{10}^2}\ \int_{\partial\,{\cal M}^\star} \ d^9 \,x \ \sqrt{-\,\tilde{g}}\ \delta N_n\  D_m\left[ \alpha^{mn} \ - \ g^{mn} \, \alpha \right] \ , \label{deltaforA}
\eeq
where
\beq
\alpha^{mn} \ = \ \frac{1}{\cal N} \ \left(D^m\,\alpha^n \ + \ D^n\,\alpha^m \right) 
\eeq
and $\alpha$ is its trace. Around the background of eqs.~\eqref{back_epos_fin2}, after an integration by parts, eq.~\eqref{deltaforA} reduces to
\beq
\delta\left(\Delta\,{\cal S}\right) \ = \ \frac{1}{2\,k_{10}^2}\ \int_{\partial\,{\cal M}^\star} \ d^9 \,x \ \frac{\sqrt{-\,\tilde{g}}}{\cal N}\ \delta g_{nr} \left[ \Box_9\,A^n \ - \ \partial^n\,\partial^m\,A_m \right] \ ,
\eeq
and thus involves the flat Maxwell kinetic operator for $A^m$, 
since in the background of eqs.~\eqref{back_epos_fin2} all ``reduced'' nine--dimensional ADM covariant derivatives $D^m$ are flat and do not involve $r$, on which ${\cal N}$ depends.
The fluctuations around the background of eqs.~\eqref{back_epos_fin2} are thus consistent with a gauge--invariant formulation for the St\"uckelberg vector field $A_m$. 

\section{\sc The Effective Orientifolds of the Background} \label{sec:background_or}

In~\cite{ms22_1} the dynamics of a probe D brane unveiled, near the $z=0$ end of the internal interval, a gravitational repulsion and a charge attraction of equal magnitude, two effects that point to an effective BPS orientifold localized there. In~\cite{ms22_1} we did not compute the tension and charge of this object, but we did show that they have opposite signs and equal magnitudes. A closer look at the background can determine them precisely, as we can now show.

As described in Appendix~\ref{app:background}, the equations solved by the background contain indeed contact terms localized at the singular ends of the interval. In particular, eqs.~\eqref{hamiltonian_F} imply for the Einstein tensor the singular limiting behavior
\beq
G_{\mu\nu} \ = \  \frac{\delta(z)}{3\,z} \ g_{\mu\nu} \ e^{- \, 2 A} \ + \ \ldots  \label{tension_id_tex}
\eeq
near the origin. Taking eqs.~\eqref {lim_eqs_z0} into account, the preceding result can be cast in the suggestive form
\beq
\sqrt{- g}\, G_{\mu\nu} \ = \   H \sqrt{- \, \det\,g_{\mu\nu}} \ g_{\mu\nu} \ \delta(z) \ + \ \ldots  \ , \label{tension_id_tex2}
\eeq
where we have added to the left--hand side the full metric determinant, which equals one in this coordinate system, to emphasize the link with the complete Einstein equations. The contact term can thus be associated to the variation of
\beq
\Delta\,{\cal S}_1 \ = \ \frac{H}{k_{10}^2} \ \int_{\cal M} d^{10}\, x \ \sqrt{-\, \det\,g_{\mu\nu}}\ \delta(z) \ . \label{deltas_1}
\eeq
thanks to the special behavior of the background metric $g_{\mu\nu}$ as $z\to 0$, which grants that
\beq
\det\left(g_{\mu\nu}\right) \ = \ e^{4A} \ \sim \ \frac{1}{3\,H\,z} \ e^{\,-\,2 A} \ .
\eeq
Note that eq.~\eqref{deltas_1} lacks the full nine--dimensional covariance, but this difficulty could be overcome considering the term in the second line of eq.~\eqref{SaSg},
\beq{}{}{}{}{}{}{}{}{}{}{}{}{}{}{}{}{}{}{}{}{}{}{}{}{}{}{}{}{}{}{}{}{}{}{}{}{}{}{}{}{}{}{}{}{}{}{}{}{}{}{}{}{}{}{}{}{}{}{}{}{}{}{}{}{}{}{}{}{}{}{}{}
\Delta\,{\cal S}_2 \ = \ \frac{1}{4!\ k_{10}^2} \ \int_{z=0} \epsilon_{\alpha_1 \ldots \alpha_4}\ e^{\alpha_1}\,\wedge\,\cdots\,e^{\alpha_4} \,\wedge \, {\cal H}_5 \ , \label{deltas_2}
\eeq
where $e^\alpha$ is the four--dimensional vielbein one-form. This alternative expression still breaks the ten--dimensional Lorentz symmetry to $SO(1,3)\times SO(5)$, just like the boundary conditions of Fermi fields in~\cite{ms21_1,mslett_20}, but possesses ten--dimensional general covariance. Eqs.~\eqref{deltas_2} and ~\eqref{deltas_1} coincide in the background, and integrating over the internal torus, whose volume shrinks to zero as $z \to 0$, turns them into the DBI action for an effective $O3$ orientifold. The resulting tension is
\beq
T \ = \ - \ \frac{\Phi}{k_{10}^2} \ , \label{tension}
\eeq
where
\beq
\Phi \ = \ H \left( 2\,\pi\,R \right)^5 
\eeq
is the five--form flux on the internal torus. The tension $T$ of the effective orientifold is indeed negative, consistently with the probe analysis in~\cite{ms22_1}.

Near the other end of the internal interval, which lies at $z=z_m$, a closer look at the background of~\eqref{back_epos_fin2} reveals the presence of additional contact terms in the Einstein equations, so that the counterpart of eq.~\eqref{tension_id_tex} reads
\bea
 G_{\mu\nu} &=&  - \ \frac{\delta\left(z_m \ - \ z\right)}{z_m \ - \ z} \ g_{\mu\nu} \ e^{-\,2 A} \ + \ \ldots \ , \nonumber \\
  G_{ij} &=&   - \ \frac{4\left(5 \ + \ \sqrt{10} \right)}{15}\ \frac{\delta\left(z_m \ - \ z\right)}{z_m \ - \ z}\ g_{ij} \ e^{-\,2 A} \ + \ \ldots \ . \label{contact_terms}
 \eea
These additional contact terms concern both $ G_{\mu\nu}$ and $ G_{ij}$, and the different limiting behavior of the metric requires a more detailed scrutiny of the available options.

The simplest alternative to eq.~\eqref{deltas_1} is provided by counterterms involving a single $z$-derivative. In addressing them, we continue to excise small regions around the ends of the interval, thus working within the regularized bulk manifold ${\cal M}^\star$.

We can now try to link the contact term in eqs.~\eqref{contact_terms} to an extrinsic curvature term,
\beq
\Delta{\cal S}_3 \ = \ \frac{\alpha}{2\,\k_{10}^2} \ \int d^9 x \ \sqrt{-\,\tilde{g}} \ \tilde{g}^{mn}\,{\cal K}_{mn} \ ,
\eeq
whose variation is
\beq
\delta\left(\Delta\,{\cal S}_3\right) \ = \ - \ \frac{\alpha}{4\,\k_{10}^2} \ \int \ d^9x \ \sqrt{-\tilde{g}} \ e^{-A} \ \left(4\,A_z \ + \ 5\,C_z\right) \left( \delta\,\tilde{g}^{\mu\nu}\ \tilde{g}_{\mu\nu} \ + \ \delta\,\tilde{g}^{ij}\ \tilde{g}_{ij}\right) \ ,
\eeq
taking eqs.~\eqref{rightend} into account. In fact, eqs.~\eqref{rightend} imply that the variations of $\tilde{g}^{\mu\nu}\,{\cal K}_{\mu\nu}$ and $\tilde{g}^{ij}\,{\cal K}_{ij}$ vanish, so that these results are only determined by the variation of the determinant. The resulting contributions to the $\mu\nu$ and $ij$ equations
are proportional to those that a nine--dimensional cosmological term would produce. 

A contribution proportional to the term in eq.~\eqref{deltas_2}, or equivalently to the term in eq.~\eqref{deltas_2},
\beq{}{}{}{}{}{}{}{}{}{}{}{}{}{}{}{}{}{}{}{}{}{}{}{}{}{}{}{}{}{}{}{}{}{}{}{}{}{}{}{}{}{}{}{}{}{}{}{}{}{}{}{}{}{}{}{}{}{}{}{}{}{}{}{}{}{}{}{}{}{}{}{}
\Delta\,{\cal S}_4 \ = \ - \ \lim_{Z^\star \to z_m} \ \frac{T(Z^\star)}{4!\ \Phi \ k_{10}^2} \ \int \epsilon_{\alpha_1 \ldots \alpha_4}\ e^{\alpha_1}\,\wedge\,\cdots\,e^{\alpha_4} \,\wedge \, {\cal H}_5 \ , \label{deltas_2m}
\eeq
is also compatible with the residual symmetry of the background, and its variation is
\beq{}{}{}{}{}{}{}{}{}{}{}{}{}{}{}{}{}{}{}{}{}{}{}{}{}{}{}{}{}{}{}{}{}{}{}{}{}{}{}{}{}{}{}{}{}{}{}{}{}{}{}{}{}{}{}{}{}{}{}{}{}{}{}{}{}{}{}{}{}{}{}{}
\delta\left(\Delta\,{\cal S}_4\right) \ = \ \lim_{Z^\star \to z_m} \ \frac{T(Z^\star)}{2\ k_{10}^2} \ \int_{z=Z^\star} \sqrt{- \ \,g_{\mu\nu}} \ \delta\,g^{\mu\nu}\, g_{\mu\nu}  \label{deltas_2m2} \ .
\eeq

Taking eqs.~\eqref{contact_terms} and the additional contributions from $\Delta\,{\cal S}_3$ and $\Delta\,{\cal S}_4$ into account, one finds that the contact terms in eqs.~\eqref{contact_terms} can be accounted for if
\beq
\alpha \ = \  \frac{8\,\sqrt{10}}{45} \left(1\ + \ \sqrt{10}\right) \ , \qquad
T(Z^\star) \ = \  - \ \frac{5 \ + \ 4\sqrt{10}}{15\left(z_m - Z^\star\right)} \ e^{- A+5 C} \ .
\eeq
Note, however, that $T(Z^\star)$ tends to $-\,\infty$ as $Z^\star$ approaches $z_m$ from lower values.

We can now turn to the simpler task of identifying the charges lying at the ends of the interval. 
To this end, we take as our starting point the naive action for the four-form gauge potential,
\beq{}{}{}
{\cal S} \ = \ - \ \frac{1}{2\times 5!\,k_{10}^2} \ \int d^{10}\,x \sqrt{-g} \ {\cal H}_{MNPQR}\ {\cal H}^{MNPQR} \ ,
\eeq
which must be supplemented by the self-duality condition. In the bulk the equation of motion is
\beq
\partial_M \left[ \sqrt{- g}\ {\cal H}^{MNPQR} \right] \ = \ 0 \ ,
\eeq
and taking into account the detailed form of the background in eqs.~\eqref{back_epos_fin2} one can see that, for $0 < z< z_m$,~\footnote{In our conventions $\epsilon_{0123}=1= - \epsilon^{0123}$, and this choice determines the overall sign in eq.~\eqref{BCHT3}.}
\beq
\sqrt{- g} \ {\cal H}^{\mu\nu\rho\sigma z}  \ = \ H \ \epsilon^{\mu\nu\rho\sigma} \ .
\eeq
With ${\cal H}$ vanishing outside the interval one can thus conclude that
\beq
\partial_z \left[ \sqrt{- g} \ {\cal H}^{\mu\nu\rho\sigma z}\right]   \ = \ H \ \epsilon^{\mu\nu\rho\sigma} \, \Big[ \delta(z) \ - \ \delta(z_m - z)\Big] \ .
\eeq 
More precisely, these results could be obtained resorting to the Henneaux-Teitelboin action~\cite{henneaux}, which can be adapted to the presence of the boundaries at the ends of the internal interval, compatibly with the full residual symmetry of the background, but the conclusion is at any rate that the bulk action should be supplemented with the contribution
\beq{}{}{}
{\cal S}_Q \ = \ - \ \frac{1}{k_{10}^2} \ \int_{\partial\,{\cal M}_{10}} {\cal B} \, \wedge\,{\cal H}_5^{(0)} \ . \label{BCHT2}
\eeq
However, writing
\beq
{\cal H}_{5} \ = \ {\cal H}_5^{(0)}{} \ + \ d\,{\cal B} \ , 
\eeq
and taking into account that
$d\left({\cal B} \wedge {\cal B}\right) = 2 \,{\cal B} \wedge d\,{\cal B}$, eq.~\eqref{BCHT2} is equivalent to the background independent expression
\beq{}{}{}
{\cal S}_Q \ = \ - \ \frac{1}{k_{10}^2} \ \int_{\partial\,{\cal M}_{10}} {\cal B} \, \wedge\,{\cal H}_{5} \ . \label{BCHT3}
\eeq
Finally, after integrating over the internal torus, one can conclude that ${\cal S}_Q$ adds to the five--dimensional effective action resulting from the compactification the terms
\beq{}{}{}{}
{\cal S}_Q \ = \ \lim_{z^\star \to 0, \, Z^\star \to z_m}\left[- \ \frac{\Phi}{k_{10}^2} \int_{{\cal M}_4\left(Z^\star\right)} {\cal B} \ + \ \frac{\Phi}{k_{10}^2} \int_{{\cal M}_4\left(z^\star\right)} {\cal B}\ \right]\ , \label{2.122}
\eeq
where $\Phi$ denotes, as in eq.~\eqref{tension}, the five--form flux over the internal torus. This expression
associates opposite five--form charges to the two endpoints of the internal interval. In particular, the contribution at the left end captures precisely the positive charge of the effective $O_3$ orientifold at $r = 0$, and comparing with eq.~\eqref{tension} finally shows that the magnitudes of the charge and tension present there are such that
\beq{}{}{}
Q \ = \ - \ T \ ,
\eeq
consistently with the analysis in~\cite{ms22_1}.

Summarizing, the low--energy five--dimensional theory for the background of eqs.~\eqref{back_epos_fin2} hosts, at one end of the interval, an effective BPS $O_3$ orientifold with negative tension and positive five--form charge, {whose opposite values are proportional to the five-form flux on the internal torus.} At the other end, the low--energy effective action contains an extrinsic curvature term and a singular tension term, but the corresponding charge is, as expected, opposite to the one present at the origin.

\section*{\sc Acknowledgments}
\vskip 12pt
We are grateful to G.~Dall'Agata, E.~Dudas and A.~Tomasiello for stimulating discussions. AS was supported in part by Scuola Normale, by INFN (IS GSS-Pi) and by the MIUR-PRIN contract 2017CC72MK\_003. JM is grateful to Scuola Normale Superiore for the kind hospitality while this work was in progress. AS is grateful to Universit\'e de Paris Cit\'e and DESY--Hamburg for the kind hospitality, and to the Alexander von Humboldt foundation for the kind and generous support, while this work was in progress. Finally, we are both very grateful to Dr.~M.~Nardelli, who kindly helped us to retrieve some mathematical literature.

\begin{appendices}
\section{\sc Some Properties of the Background} \label{app:background}
In this Appendix we collect some useful properties of the background of eqs.~\eqref{back_epos_fin2} of Section~\ref{sec:intro}. 
In our conventions, capital Latin labels like $M$ denote curved ten--dimensional indices, while Greek or Latin labels like $(\mu,r,i)$ denote their spacetime or internal portions. 
Moreover, when we need to distinguish the curved radial index $r$ or $z$ from the remaining nine--dimensional ones, as in the ADM decomposition of Section~\ref{sec:gen_rel_boundary}, we denote them collectively by $m$. 
We use a ``mostly--plus'' signature, defining the Riemann curvature tensor via~\footnote{These conventions are as in~\cite{ms21_1} and \cite{ms21_2}.}
\beq
[ \nabla_M, \nabla_N] V_P \ = \ {R_{MNP}}^{Q}\, V_Q \ ,
\eeq
so that
\beq
{R_{MNP}}^{Q} \ = \ \partial_N\,{\Gamma^Q}_{MP} \ - \ \partial_M\,{\Gamma^Q}_{NP}  \ + \ {\Gamma^Q}_{NR}\, {\Gamma^R}_{MP} \ - \ {\Gamma^Q}_{MR}\, {\Gamma^R}_{NP} \ ,
\eeq
and define the Ricci tensor as
\beq
R_{MQ} \ = \ {R_{MNQ}}^{N} \ .
\eeq

The background~\eqref{back_epos_fin2} is of the type
\beq
ds^2 \ = \ e^{\,2\,A(r)}\, dx^2 \ + \ e^{\,2\,B(r)}\, dr^2 \ + \ e^{\,2\,C(r)}\, dy^2 \ , \label{metricABCapp}
\eeq
where the $x^\mu$--coordinates, with $\mu=0,\ldots,3$, refer to the four--dimensional spacetime, while the $y^i$--coordinates, with $i=1,\ldots,5$ refer to the internal torus. We work mostly in terms of the $z$ variable, defined by
\beq
z \ = \ \int_0^r d\xi \ e^{B(\xi)-A(\xi)} \ ,
\eeq
whose upper limit is
\beq
z_m \ = \ \int_0^\infty d\xi \ e^{B(\xi)-A(\xi)} \ \simeq \ 2.24\ z_0 \ , \label{zm_app}
\eeq
where
\beq{}{}{}{}{}{}{}{}{}{}{}{}{}{}{}{}{}{}{}{}
z_0  \ = \ \rho\, h^\frac{1}{2} \ = \ \left(2 H \rho^3\right)^\frac{1}{2} \ , \qquad \mathrm{with} \qquad  h \ = \ 2\,H\,\rho \ . \label{z0_iap}
\eeq

When the Einstein equations
\beq
G_{MN}  =  \frac{1}{4!}\ \left({\cal H}_5{}^2\right)_{M N} = \frac{1}{24} \ g^{PP'}\,g^{QQ'}\,g^{RR'}\,g^{SS'}\ {\cal H}_{5\,M P Q R S}\ {\cal H}_{5\,N P' Q' R' S'} 
\eeq
are written for the background~\eqref{metricABCapp}, in the $z$ coordinate the $G^{(0)}{}_{zz}$ equation, which we often refer to as Hamiltonian constraint, becomes
\beq
 3\left(A_z\right)^2 \ + \ 10\, A_z\, C_z \ + \ 5 \left(C_z\right)^2 \ = \ - \ 2\ {\cal W}_5^2 \ ,
\eeq
where we have introduced the convenient combination
\beq{}{}
{\cal W}_5 \ = \ \frac{h}{4\,\rho}\ e^{A-5C} \ , 
\eeq
while the remaining Einstein equations take the form 
\bea{}{}{}{}{}{}{}{}{}{}{}{}{}{}{}{}{}{}
G_{\mu\nu} \!\!&\equiv&\!\! g_{\mu\nu} e^{-\,2 A}\,\left[3\,A_{zz} \ +\  5\,C_{zz}\ + \ \left(3 A_z+5 C_z\right)^2 \ + \ 4\,{\cal W}_5{}^2  \right] \ = \ - \ 4\,{\cal W}_5^2 \ e^{\,-\,2 A} \ g_{\mu\nu}\, , \label{Gmnij} \\
G_{ij} \!\!&\equiv&\!\! g_{ij}e^{-\, 2 A}\,\left[4\left(A_{zz} \ +\  C_{zz}\right)\ + \ 4 \left(3 A_z+5 C_z\right)\left(A_z + C_z\right) \ + \ 4\,{\cal W}_5{}^2  \right] \, = \,  4\,{\cal W}_5^2 \ e^{\,-\,2 A} g_{ij}\, . \nonumber 
\eea
Combining them, one can deduce that
\beq
A_{zz} \ = \ 4\,{\cal W}_5^2 \ -  \ \left(3 A_z + 5 C_z\right )A_z \ , \qquad C_{zz} \ = \ - \ 4\,{\cal W}_5^2 \ - \ \left(3 A_z + 5 C_z\right)C_z\ . \label{hamiltonian_F}
\eeq
Recalling also eqs.~\eqref{back_epos_fin2}, close to $z=0$
\beq
\frac{z}{z_0} \ \sim \  \frac{2}{3}\left(\frac{r}{\rho}\right)^\frac{3}{2}  \ , \qquad
\frac{r}{\rho} \ \sim \ \left(\frac{3 \,z}{2\,z_0}\right)^\frac{2}{3} \ ,
\eeq
and therefore
\bea
e^A &\sim& \frac{1}{h^\frac{1}{4}}\left(\frac{2 \,z_0}{3 \,z}\right)^\frac{1}{6} \, \qquad
e^C \ \sim \ h^\frac{1}{4}\left(\frac{3 \,z}{2 \,z_0}\right)^\frac{1}{6} \,, \nonumber \\
A_z &\sim& - \ \frac{1}{6\,z} \ , \qquad
C_z \ \sim \  \frac{1}{6\,z} \ , \qquad {\cal W}_5^2 \ \sim \ \frac{1}{36\,z^2} \ . \label{origin}
    \eea
    The leading behavior of the metric and five--form backgrounds close to $z=0$ is thus
\bea
&& ds^2 \ \sim \ \frac{dx^2 \ + \ dz^2}{\left(3\,\left|H\right|z\right)^\frac{1}{3}} \ + \ \left(3\,\left|H\right|z\right)^\frac{1}{3}\ d\,\vec{y}^{\,2} \ , \nonumber \\
&& {\cal H}_5 \ \sim \ H \left\{ \frac{dx^0 \wedge ...\wedge dx^3\wedge dz}{\left[3\left|H\right|z\right]^\frac{5}{3}} \ + \ dy^1 \wedge ... \wedge dy^5\right\} \ , \label{lim_eqs_z0}
\eea
but the limiting behavior of $A_{zz}$ and $C_{zz}$ actually includes contact terms
\beq
z\, A_{zz} \ \sim \ \frac{1}{6\,z} \ - \ \frac{1}{6} \ \delta(z) \ , \qquad z\, C_{zz} \ \sim \ - \ \frac{1}{6\,z} \ + \ \frac{1}{6} \ \delta(z) \ .
\eeq
This can be seen, for example, if one regulates the two limiting forms for $A_z$ and $C_z$ according to
\beq
A_z \ \sim \ - \ \lim_{\epsilon \to 0} \ \frac{1}{6\,\sqrt{z^2 + \epsilon^2}} \ , \qquad
C_z \ \sim \ \lim_{\epsilon \to 0} \  \frac{1}{6\,\sqrt{z^2 + \epsilon^2}} \ .
\eeq

As a result, the contact terms are to be included in eqs.~\eqref{hamiltonian_F}, which become
\bea
&& z \ A_{zz} \ = \ 4\,z\, {\cal W}_5^2 \ -  \ z \left(3 A_z + 5 C_z\right )A_z \ - \ \frac{1}{6}\ \delta(z) \ , \nonumber \\
&& z\ C_{zz} \ = \ - \ 4\,z\, {\cal W}_5^2 \ - \ z \left(3 A_z + 5 C_z\right)C_z  \ + \ \frac{1}{6}\ \delta(z)\ , \label{hamiltonian_F2}
\eea
and consequently one can conclude that the equation for $g_{\mu\nu}$ in~\eqref{Gmnij} is actually
\beq
 G_{\mu\nu} \ = \  \frac{\delta(z)}{3\,z} \ g_{\mu\nu} \ e^{-\,2 A} \ + \ \ldots \ , \label{tension_id}
\eeq
and includes a contact term at the origin, while the equation for $g_{ij}$ does not.

As $z$ approaches the finite value $z_m$ of eq.~\eqref{zm_app}, one can see that
\bea{}{}
&& \frac{z_m - z}{z_0} \,  \sim \,
\frac{\sqrt{2}}{3}\left(\sqrt{10}\,+\,2\right)\  e^{\,-\,\frac{r}{4\rho}\left(\sqrt{10}\,-\,2\right)} \ , \qquad
e^{\,-\,\frac{r}{2\,\rho}} \,\sim\, \left[\frac{\sqrt{5}-\sqrt{2}}{2} \left(\frac{z_m-z}{z_0}\right)\right]^{\frac{\sqrt{10}\,+\,2}{3}}  \,, \nonumber \\
&& e^{2A} \ \sim \ \sqrt{\frac{2}{h}} \left[\frac{\sqrt{5}-\sqrt{2}}{2} \left(\frac{z_m-z}{z_0}\right)\right]^{\frac{\sqrt{10}\,+\,2}{3}}  \,, \qquad
e^{2C} \! \sim \! \sqrt{\frac{h}{2}}\left[\frac{\sqrt{5}-\sqrt{2}}{2} \left(\frac{z_m-z}{z_0}\right)\right]^{\,-\,\frac{\sqrt{10}}{5}} , \nonumber \\
&& A_z  \ \sim \ {-} \ \frac{1}{6} \ \frac{\sqrt{10}\,+\,2}{z_m - z} \ , \qquad\qquad
C_z \, \sim  \ \frac{1}{\sqrt{10}} \ \frac{1}{z_m - z} \ ,\nonumber \\
&& {\cal W}_5 \, \sim \,  \frac{\sqrt{2}}{2\,z_0}   \left[\frac{\sqrt{5}\,-\,\sqrt{2}}{2} \left(\frac{z_m-z}{z_0}\right)\right]^{\frac{2 \sqrt{10}\,+\,1}{3}}\ .  \label{large_rn}
\eea

The leading behavior of the metric and five--form backgrounds close to $z=z_m$ is thus
\bea
&& ds^2 \sim \sqrt{\frac{2}{h}} \left[\frac{\sqrt{5}-\sqrt{2}}{2} \left(\frac{z_m-z}{z_0}\right)\right]^{\frac{\sqrt{10}\,+\,2}{3}} \!\!\!\!\!\!\!\!  \left( dx^2 + dz^2\right) \, + \, \sqrt{\frac{h}{2}}\left[\frac{\sqrt{5}-\sqrt{2}}{2} \left(\frac{z_m-z}{z_0}\right)\right]^{\,-\,\frac{\sqrt{10}}{5}} \!\!\!\!\!\!\!\!\!\! d\,\vec{y}^{\,2} , \nonumber \\
&& {\cal H}_5 \sim H \left\{ \left(\frac{2}{h}\right)^\frac{5}{2} \left[\frac{\sqrt{5}-\sqrt{2}}{2} \left(\frac{z_m-z}{z_0}\right)\right]^{\frac{4\,\sqrt{10}\,+\,5}{3}}\!\!\!\!\!\!\!\!\!\!\!\! dx^0 \wedge ...\wedge dx^3\wedge dz \, + \, dy^1 \wedge ... \wedge dy^5\right\} . 
\eea

From these expressions, proceeding as above, one can deduce that
\bea
\left(z_m \ - \ z\right) A_{zz} &=&  - \  \frac{\sqrt{10}\,+\,2}{6} \ \delta\left(z_m \ - \ z\right) \ + \ \ldots \ , \nonumber \\
\left(z_m \ - \ z\right) C_{zz} &=&   \frac{1}{\sqrt{10}}  \ \delta\left(z_m \ - \ z\right) \ + \ \ldots \ ,
\eea
and consequently the Einstein equations for $g_{\mu\nu}$ and $g_{ij}$ contain at the right end of the interval the contact terms
\bea
 G_{\mu\nu} &=&  - \ \frac{\delta\left(z_m \ - \ z\right)}{z_m \ - \ z} \ g_{\mu\nu} \ e^{-\,2 A} \ + \ \ldots \ , \nonumber \\
  G_{ij} &=&   - \ \frac{4\left(5 \ + \ \sqrt{10} \right)}{15}\ \frac{\delta\left(z_m \ - \ z\right)}{z_m \ - \ z}\ g_{ij} \ e^{-\,2 A} \ + \ \ldots \ .
 \eea
\end{appendices}

\end{document}